\documentclass{aastex61}

\date{\today}

\begin{document}

\title{Planck 2015 constraints on spatially-flat dynamical dark energy models}

\author{Junpei Ooba}
\altaffiliation{ooba.jiyunpei@f.mbox.nagoya-u.ac.jp}
\affiliation{Department of Physics and Astrophysics, Nagoya University, Nagoya 464-8602, Japan}

\author{Bharat Ratra}
\affiliation{Department of Physics, Kansas State University, 116 Cardwell Hall, Manhattan, KS 66506, USA}

\author{Naoshi Sugiyama}
\affiliation{Department of Physics and Astrophysics, Nagoya University, Nagoya 464-8602, Japan}
\affiliation{Kobayashi-Maskawa Institute for the Origin of Particles and the Universe, Nagoya University, Nagoya, 464-8602, Japan}
\affiliation{Kavli Institute for the Physics and Mathematics of the Universe (Kavli IPMU), The University of Tokyo, Chiba 277-8582, Japan}



\begin{abstract}

We determine constraints on spatially-flat tilted dynamical dark energy 
XCDM and $\phi$CDM inflation models by analyzing Planck 2015 cosmic 
microwave background (CMB) anisotropy data and baryon acoustic oscillation 
(BAO) distance measurements. XCDM is a simple and widely used but physically  
inconsistent parameterization of dynamical dark energy, while 
the $\phi$CDM model is a physically consistent one in which a scalar 
field $\phi$ with an inverse power-law potential energy density powers 
the currently accelerating cosmological expansion. Both these models 
have one additional parameter compared to standard $\Lambda$CDM and both
better fit the TT + lowP + lensing + BAO data than does the
standard tilted flat-$\Lambda$CDM model, with $\Delta \chi^2 = -1.26\ 
(-1.60)$ for the XCDM ($\phi$CDM) model relative to the $\Lambda$CDM model.
While this is a 1.1$\sigma$ (1.3$\sigma$) improvement over standard
$\Lambda$CDM and so not significant, dynamical dark energy models 
cannot be ruled out. In addition, both dynamical dark energy models 
reduce the tension between the Planck 2015 CMB anisotropy and the weak 
lensing $\sigma_8$ constraints.
\end{abstract}




\section{Introduction} \label{sec:intro}

The standard cosmological model, spatially-flat $\Lambda$CDM 
\citep{Peebles1984}, is parameterized by six cosmological parameters
conventionally taken to be:
$\Omega_{\rm b} h^2$ and $\Omega_{\rm c} h^2$, the current values
of the baryonic and cold dark matter (CDM) density parameters
multiplied by $h^2$ [where $h = H_0/ (100\ {\rm km}\ {\rm s}^{-1}\ {\rm Mpc}^{-1})$ and $H_0$ is the Hubble constant]; $\theta$, the angular diameter 
distance as 
a multiple of the sound horizon at recombination; $\tau$, the reionization 
optical depth; and $A_{\rm s}$ and $n_{\rm s}$, the amplitude and spectral 
index of the (assumed) power-law primordial scalar energy density 
inhomogeneity power spectrum \citep{PlanckCollaboration2016}. In this 
model, the currently accelerating cosmological expansion is powered by the 
cosmological constant $\Lambda$ which is equivalent to a dark energy ideal 
fluid with equation of state parameter $w_0 = -1$. For reviews of this 
model see \citet{RatraVogeley2008}, \citet{Martin2012}, and \citet{Brax2018}. 
This model assumes flat spatial hypersurfaces, which is largely consistent 
with most available observational constraints 
\citep[][and references therein]{PlanckCollaboration2016}.\footnote{Using a 
physically consistent non-flat inflation model \citep{Gott1982, Hawking1984, Ratra1985} power spectrum of energy density inhomogeneities \citep{RatraPeebles1995, Ratra2017} to analyse the Planck 2015 cosmic microwave background 
(CMB) anisotropy measurements \citep{PlanckCollaboration2016}, \citet{Oobaetal2018a} find that these data do not require flat spatial hypersurfaces in the six 
parameter non-flat $\Lambda$CDM model \citep[also see][]{ParkRatra2018a, ParkRatra2018b, ParkRatra2019d}. 
In the non-flat $\Lambda$CDM model, compared to the standard flat-$\Lambda$CDM 
model, there is no simple tilt option so $n_{\rm s}$ is no longer a free 
parameter 
and it is instead replaced by the current value of the spatial curvature 
energy density parameter $\Omega_{\rm k}$. CMB anisotropy data also do not 
require flat spatial hypersurfaces in the seven parameter non-flat XCDM 
and $\phi$CDM inflation models \citep{Oobaetal2018b, Oobaetal2018c, ParkRatra2018b, ParkRatra2018c, ParkRatra2019d}.
In both these models $n_{\rm s}$ is again replaced by $\Omega_{\rm k}$. These 
models differ from the seven parameter spatially-flat XCDM and $\phi$CDM 
inflation models we study in this paper, in which $n_{\rm s}$ is a parameter but 
$\Omega_{\rm k}$ is not.}

However, there also are suggestions that flat-$\Lambda$CDM might not be 
as compatible with different or larger compilations of cosmological 
measurements \citep{Sahnietal2014, Dingetal2015, Solaetal2015, Zhengetal2016, Solaetal2017a, Solaetal2018, Solaetal2017b, Zhaoetal2017, Solaetal2017c, Zhangetal2017, Solaetal2017d, GomezValentSola2017, Caoetal2018}
that might be more consistent with dynamical dark energy 
models.\footnote{Amongst these analyses that also make use of CMB anisotropy
data, those that have used a physically consistent dynamical dark energy
model such as $\phi$CDM \citep{Solaetal2017b, Solaetal2017c, Solaetal2017d, GomezValentSola2017} have performed only an approximate CMB anisotropy analysis.} 
The simplest, but physically inconsistent and widely used, dynamical dark 
energy parameterization is the seven parameter XCDM model in which the 
equation of state relating the pressure and energy density of the dark energy 
fluid is $p_X = w_0 \rho_X$ and $w_0$ is the additional, seventh, parameter.
The simplest physically consistent dynamical dark energy model is the seven 
parameter $\phi$CDM model, in which a scalar field $\phi$ with potential 
energy density $V(\phi) \propto \phi^{-\alpha}$ is the dynamical dark energy 
\citep{PeeblesRatra1988, RatraPeebles1988} and $\alpha > 0$ is the seventh 
parameter that governs dark energy evolution.\footnote{While
XCDM is often used to model dynamical dark energy, it is not a physically 
consistent model as it cannot describe the evolution of energy density 
inhomogeneities. Also, XCDM does not accurately model $\phi$CDM dark energy
dynamics \citep{PodariuRatra2001}.} 
In this paper we use the Planck 2015 CMB anisotropy data to constrain the 
seven parameter spatially-flat XCDM and $\phi$CDM models. \citet{Oobaetal2018c}
were the first to derive proper (non-approximate) CMB anisotropy data 
constraints on the physically consistent (non-flat) dynamical dark energy 
$\phi$CDM model.\footnote{Aside 
from CMB anisotropy measurements, many other observations have been used to 
constrain the $\phi$CDM model \citep[see, e.g.,][]{ChenRatra2004, Samushiaetal2007, Yasharetal2009, SamushiaRatra2010, ChenRatra2011b, FarooqRatra2013, Pavlovetal2014, Avsajanishvilietal2015, Farooqetal2017, Solaetal2017b, Solaetal2017c, Solaetal2017d, Zhaietal2017, GomezValentSola2017, Avsajanishvilietal2017, Ryanetal2018, Ryanetal2019, ParkRatra2018d, ParkRatra2019d, KhadkaRatra2019}.}   
In this paper 
we present results from the first complete (non-approximate) analyses of CMB 
anisotropy data using the spatially-flat tilted $\phi$CDM model.

The structure of our paper is as follows. In Sec.\ II we summarize the methods
we use in our analyses here. Our parameter constraints are tabulated, plotted, 
and discussed in Sec.\ III, where we also comment on the goodness-of-fit of 
the best-fit XCDM and $\phi$CDM models. We conclude in Sec.\ IV.

\section{Methods}

In the XCDM parameterization the equation of state of the dark energy fluid 
is $p_X = w_0 \rho_X$.
In this parameterization, to render it physically sensible, we make the 
additional (somewhat arbitrary) assumption that spatial inhomogeneities in 
the dark energy fluid propagate at the speed of light.

In the $\phi$CDM model the equations of motion are
\begin{eqnarray}
\label{eq:eom}
& \ddot\phi + 3 {\dot a \over a} \dot\phi - \kappa \alpha m_P^2 
  \phi^{-(\alpha + 1)} = 0, \\
& \left({\dot a \over a}\right)^2 = {8 \pi \over 3 m_P^2}( \rho + \rho_\phi), \\
& \rho_\phi = {m_P^2 \over 32 \pi} \left( \dot\phi^2 + 2 \kappa m_P^2  
  \phi^{-\alpha}\right) .
\end{eqnarray}
Here the scalar field potential energy density 
$V(\phi) = \kappa m_P^2 \phi^{-\alpha}$, $m_p$ is the Planck mass, and 
$\kappa$ is determined in terms of the other parameters. $a$ is the 
cosmological scale factor and an overdot represents a derivative with 
respect to time. \newpage \noindent $\rho$ and $\rho_\phi$ are the energy densities 
excluding the scalar field and that of the scalar field, respectively.
The $\phi$CDM model equations of motion has a time-dependent attractor or
tracker solution and so predictions in this model do not depend on 
initial conditions \citep{PeeblesRatra1988,RatraPeebles1988,Pavlovetal2013}.   
On this solution, the initially subdominant scalar field energy density 
evolves in a manner to attempt to become the dominant energy density; this 
mechanism could partially alleviate the fine-tuning associated with the 
currently accelerating cosmological expansion. 

\begin{figure}[ht]
\plottwo{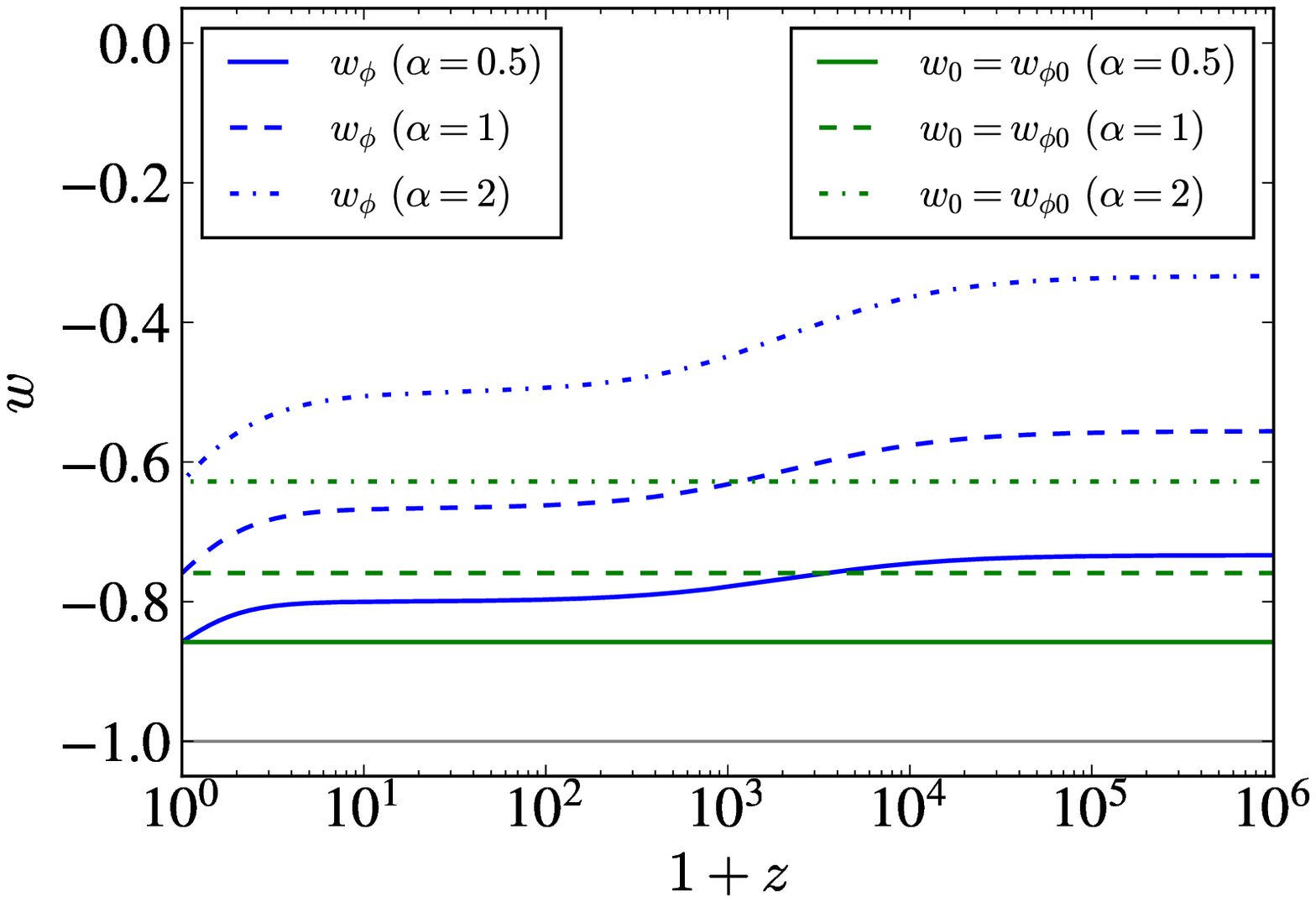}{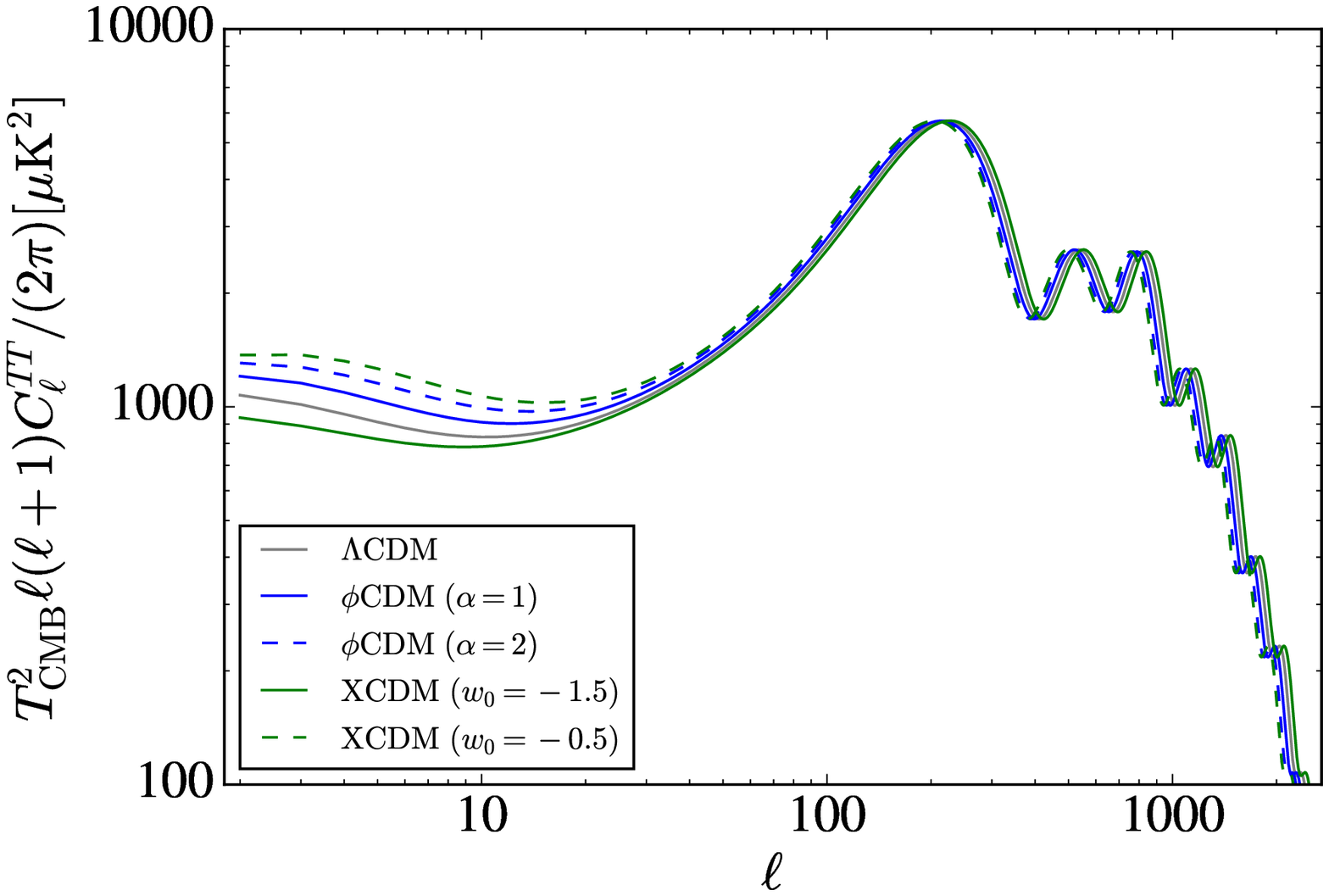}
\caption{
Left panel: Dynamical evolution of the equation of state parameter of dark 
energy in a few $\phi{\rm CDM}$ (blue) and corresponding XCDM (green) models. 
Here $w_0$ of each corresponding XCDM model is set to the present value 
of $w_{\phi}$ of each $\phi{\rm CDM}$ model. The $\Lambda{\rm CDM}$ model is 
shown as a gray solid line. Right panel: The CMB temperature anisotropy 
spectrum of two $\phi{\rm CDM}$ models (blue lines) and of two 
XCDM models (green lines). The best-fit $\Lambda{\rm CDM}$ model is shown 
as a gray solid line.
Here other cosmological parameters are fixed to the best-fit $\Lambda{\rm CDM}$
model values for the Planck 2015 TT + lowP data.
\label{fig:w_cl}
}
\end{figure}

Figure \ref{fig:w_cl} shows the dynamical evolution of the equation of state 
parameter (the ratio of pressure to energy density) of dark energy in 
some $\phi{\rm CDM}$ and XCDM models and the effects of dynamical dark energy 
on the CMB temperature anisotropy spectrum.

In this study we compute the angular power spectra of the CMB anisotropy by 
using CLASS \citep{Blasetal2011}\footnote{Our flat space $\phi$CDM CMB 
anisotropy angular power spectra differ somewhat from earlier results in 
\citet{Braxetal2000} and \citet{Mukherjeeetal2003}. We have verified that 
our results are accurate.}
and perform the Markov chain Monte Carlo analyses with Monte Python 
\citep{Audrenetal2013}. In both spatially-flat dynamical dark energy 
models the primordial power spectrum of energy density inhomogeneities 
is taken to be that generated by quantum-mechanical fluctuations in the
spatially-flat tilted inflation model \citep{LucchinMatarrese1985, Ratra1992, Ratra1989}
\begin{equation}
   P(k)=A_{\rm s} \left(\frac{k}{k_0} \right)^{n_{\rm s}},
\end{equation}
where $k$ is wavenumber and $A_{\rm s}$ is the amplitude at the pivot scale 
$k_0=0.05~\textrm{Mpc}^{-1}$.

We consider a flat prior with the ranges of the cosmological parameters 
chosen to be 
\begin{eqnarray}
\label{eq:prior}
&100\theta \in (0.5,10),\ \ \Omega_{\rm b}h^2 \in (0.005,0.04),\ \ \Omega_{\rm c}h^2 \in (0.01,0.5), \nonumber\\
&\tau \in (0.005,0.5),\ \ {\rm ln}(10^{10}A_{\rm s}) \in (0.5,10),\ \ n_{\rm s} \in (0.5, 1.5),
\end{eqnarray}
while the parameters characterizing the dark energy dynamics range over
\begin{equation}
\label{eq:prior2}
w_0 \in (-3, 0.2),\ \ \alpha \in (0, 8).
\end{equation}
The CMB temperature and the effective number of neutrinos were set to 
$T_{\rm CMB}= 2.7255\ \rm K$ from COBE \citep{Fixsen2009} and $N_{\rm eff}=3.046$ 
with one massive (0.06 eV) and two massless neutrino species in a normal 
hierarchy. The primordial helium fraction $Y_{\rm He}$ is 
inferred from standard Big Bang nucleosynthesis, as a function of the baryon 
density.

We constrain model parameters by comparing our results to the CMB angular 
power spectrum data from the Planck 2015 release 
\citep{PlanckCollaboration2016} and the baryon acoustic oscillation (BAO) 
distance measurements from the 
matter power spectra obtained by the 6dF Galaxy Survey 
\citep{Beutleretal2011}, the Baryon Oscillation Spectroscopic Survey 
(LOWZ and CMASS) \citep{Andersonetal2014}, and the Sloan Digital Sky 
Survey main galaxy sample (MGS) \citep{Rossetal2015}.

\section{Results}

\begin{figure}[ht]
\plottwo{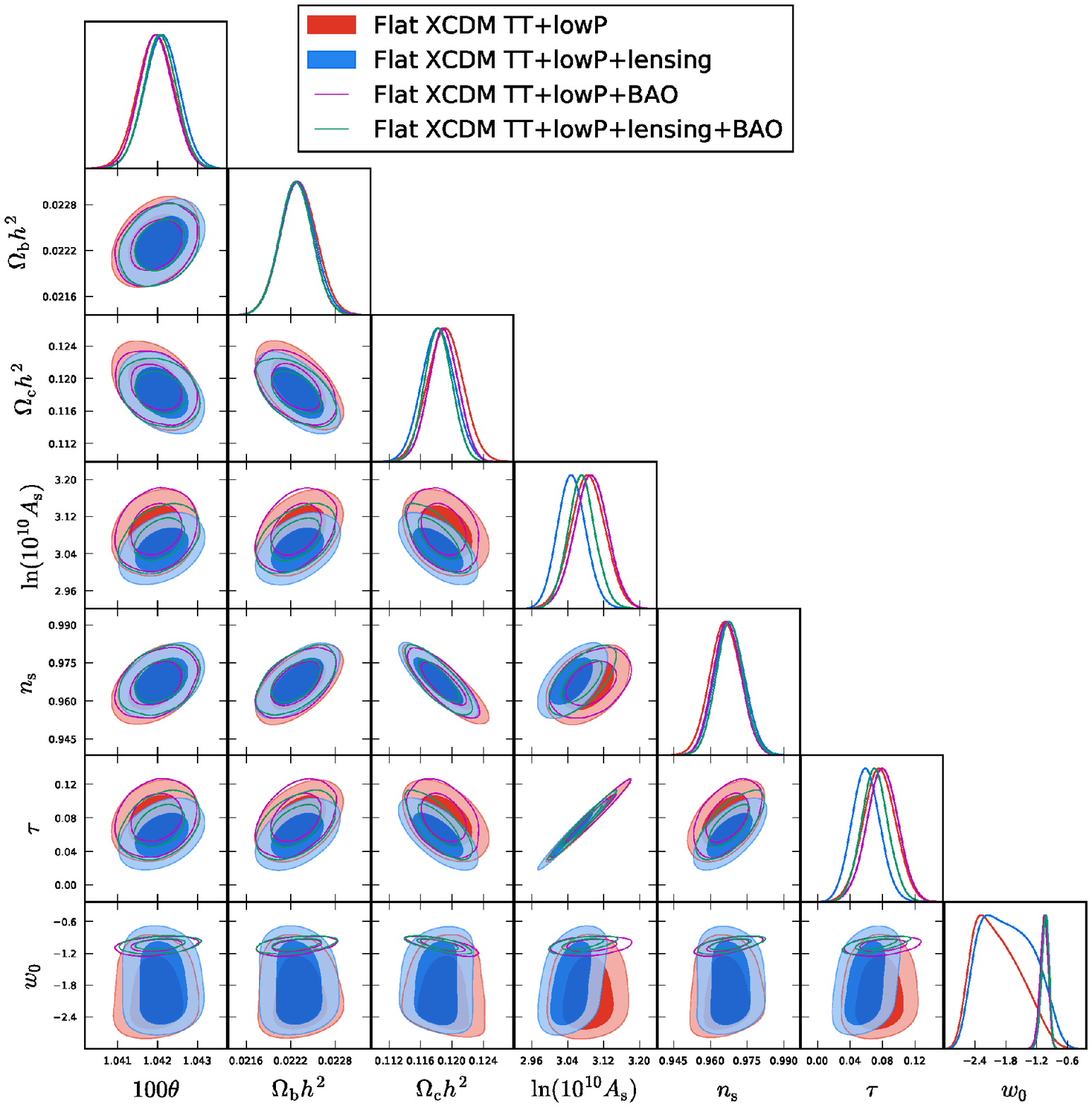}{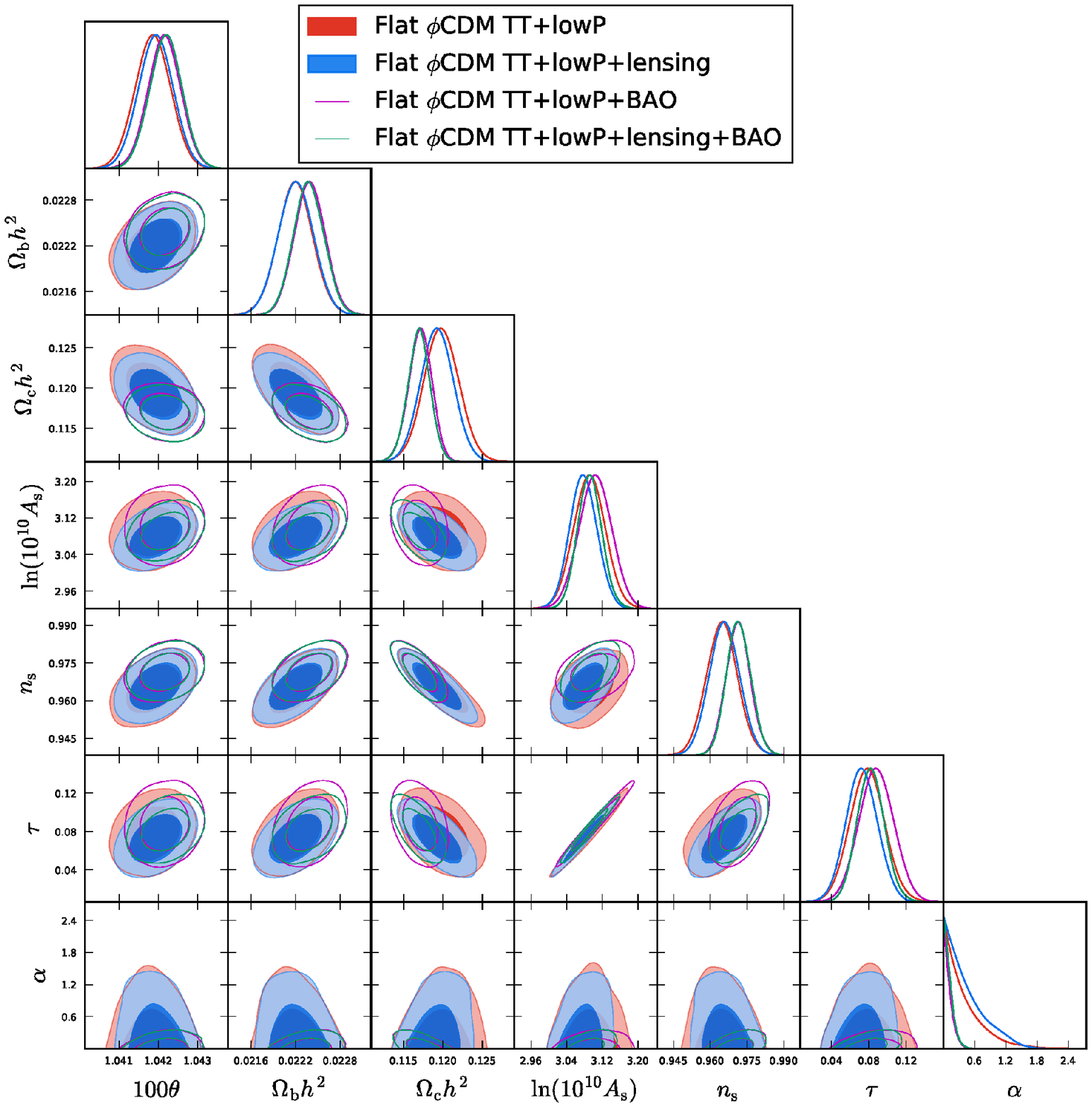}
\caption{$68.27\%$ and $95.45\%$ confidence level contours for the XCDM and
$\phi$CDM models using various data sets, with the other parameters marginalized.\label{fig:tri}}
\end{figure}

\begin{figure}[ht]
\gridline{\fig{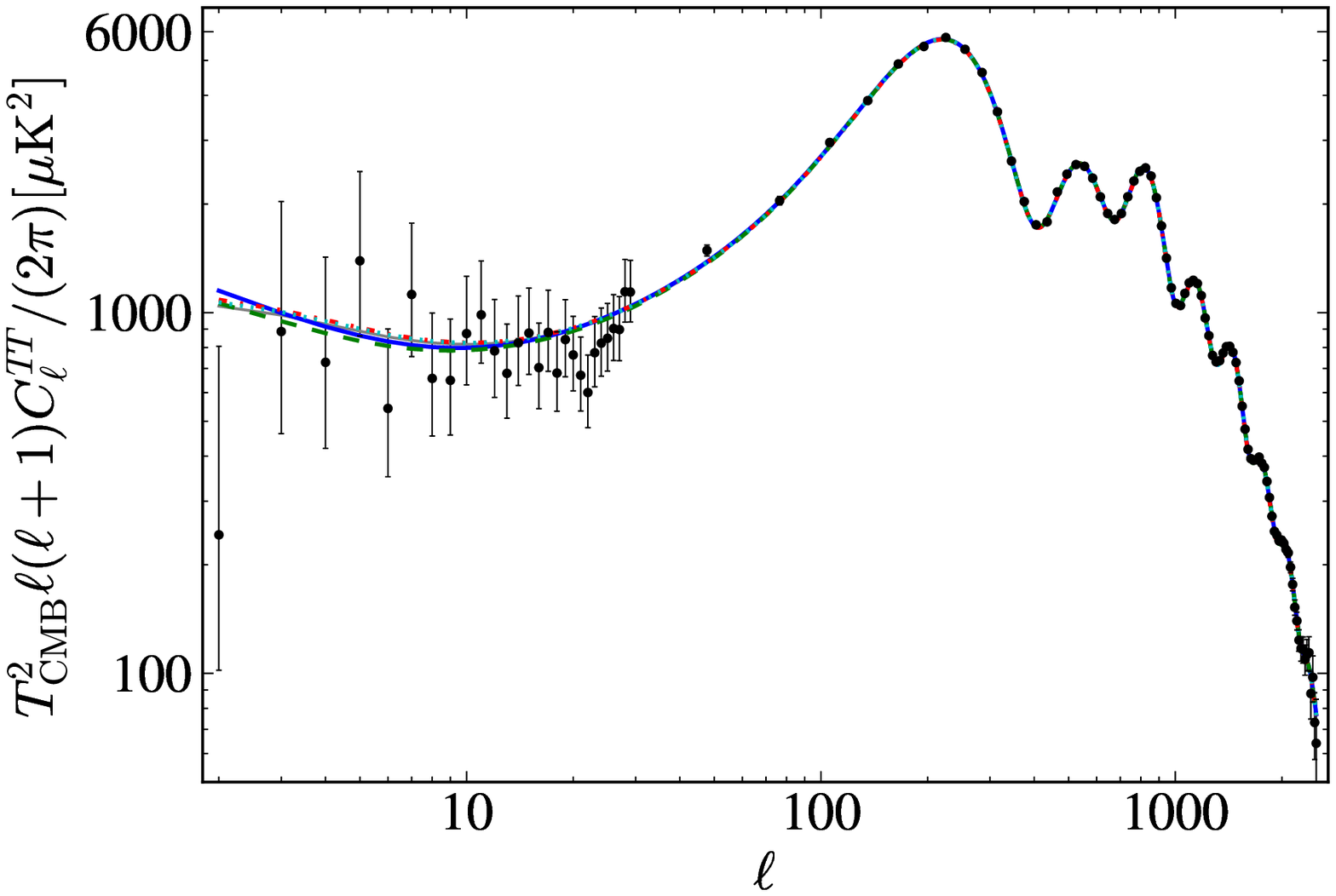}{0.45\textwidth}{(a)}
          \fig{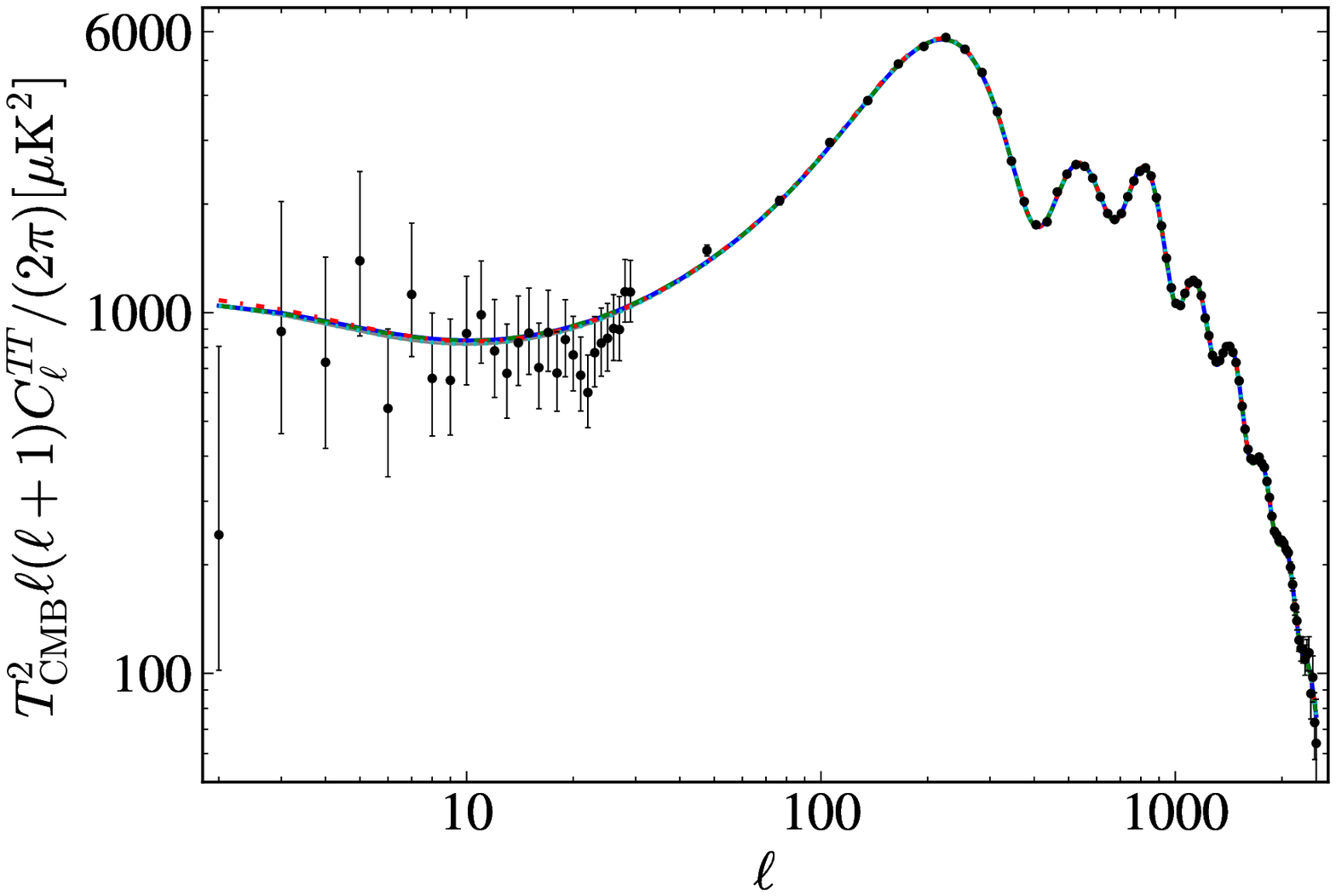}{0.45\textwidth}{(b)}
          }
\gridline{\fig{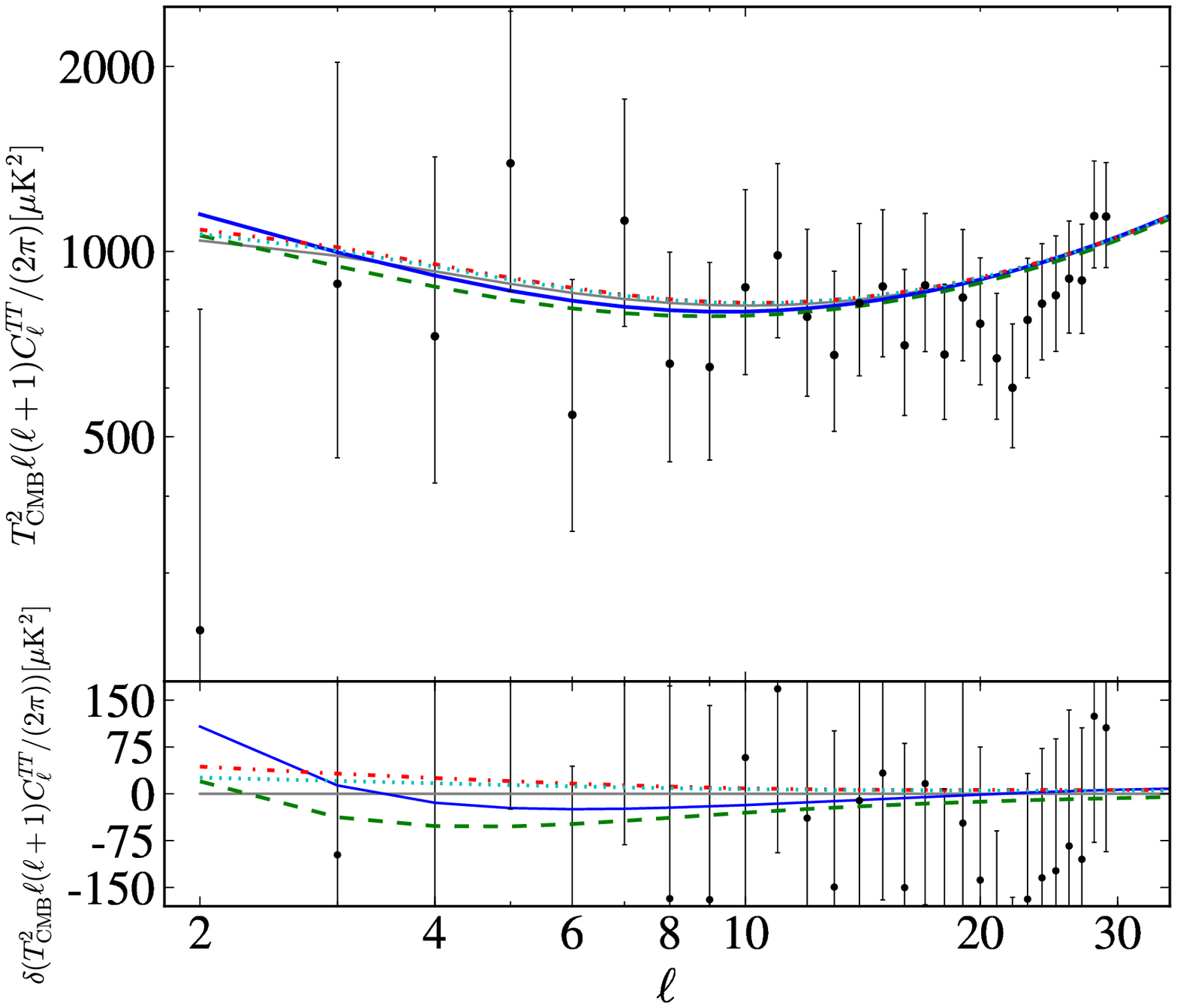}{0.42\textwidth}{(c)}
          \fig{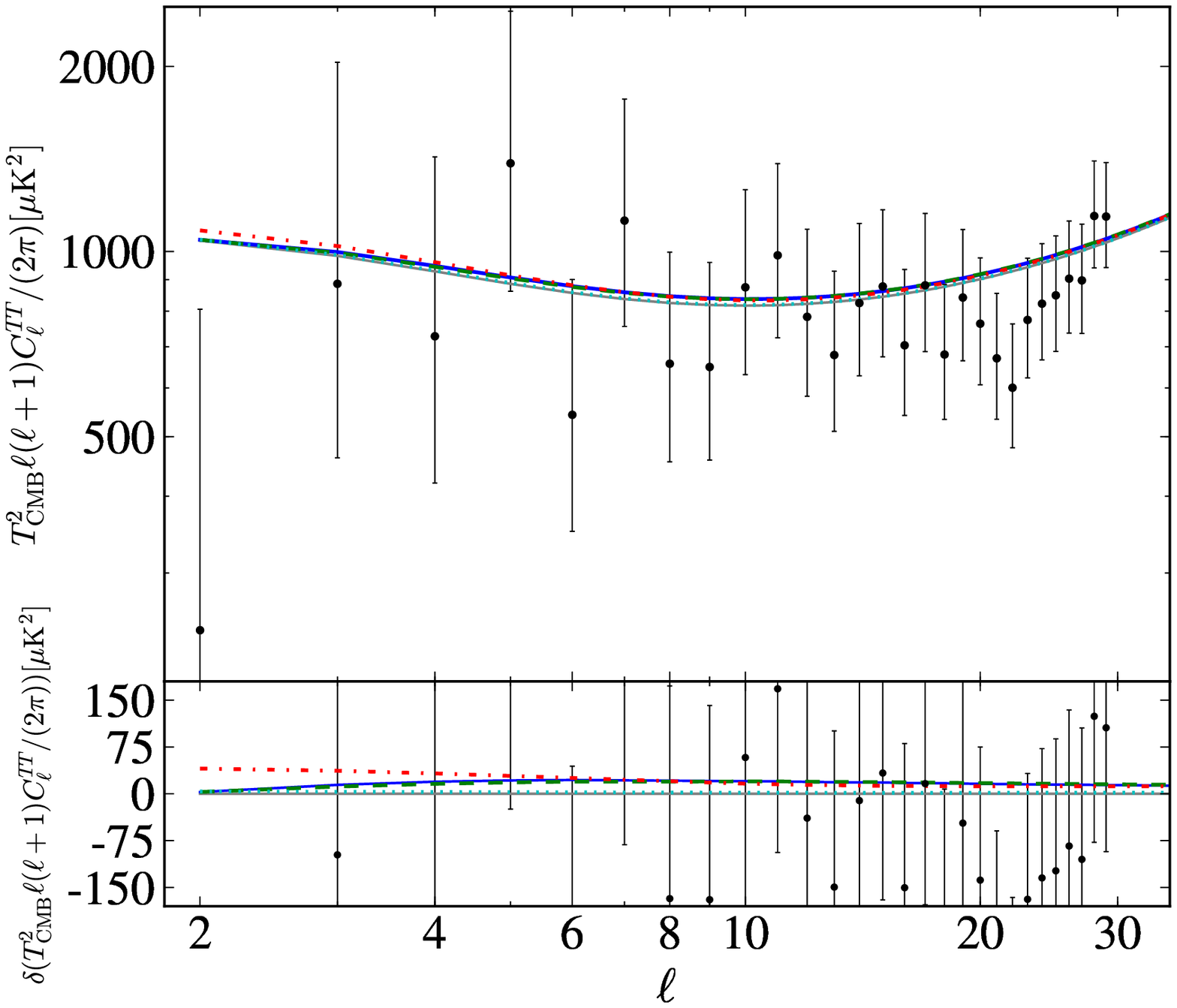}{0.42\textwidth}{(d)}
          }
\gridline{\fig{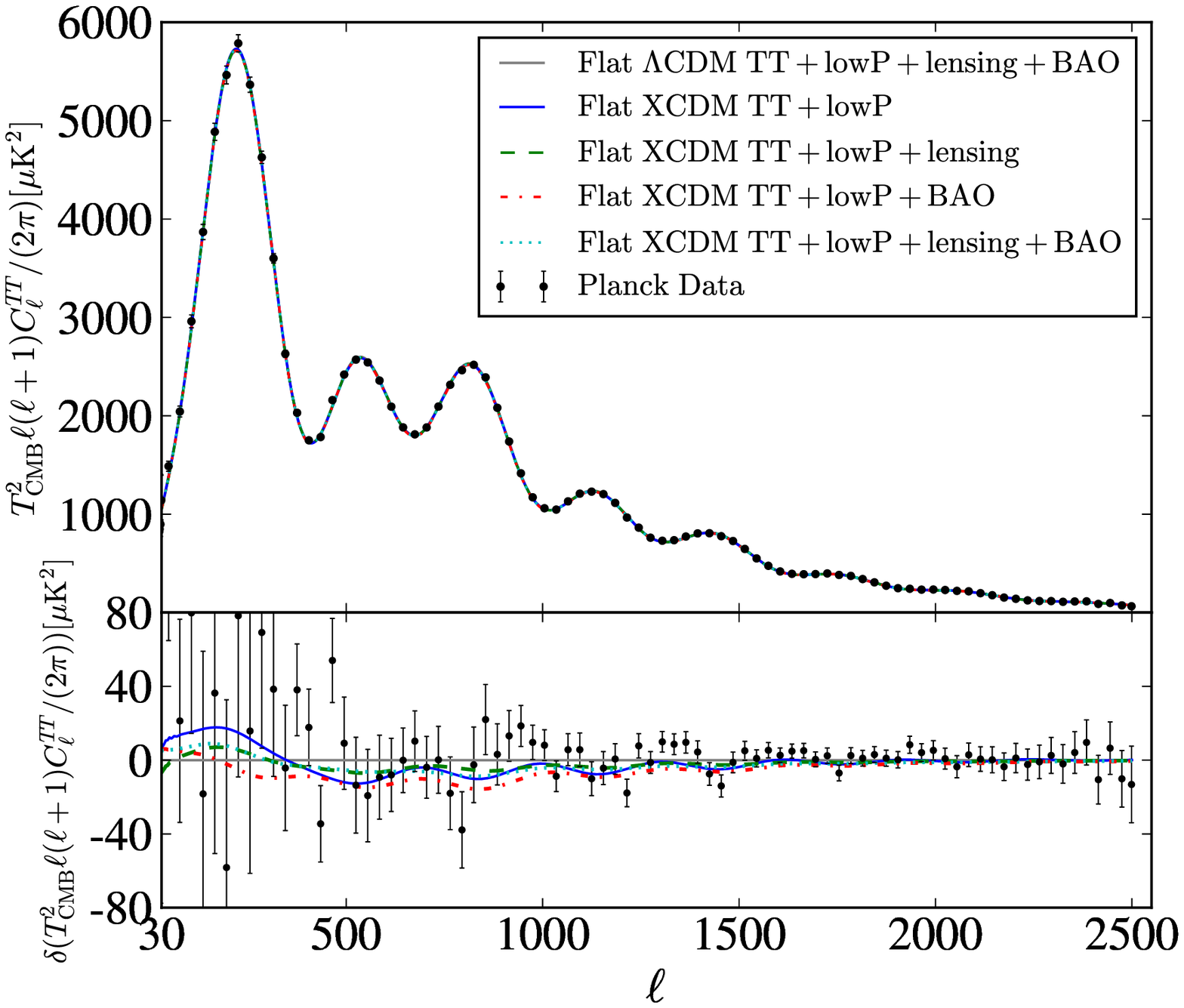}{0.43\textwidth}{(e)}
          \fig{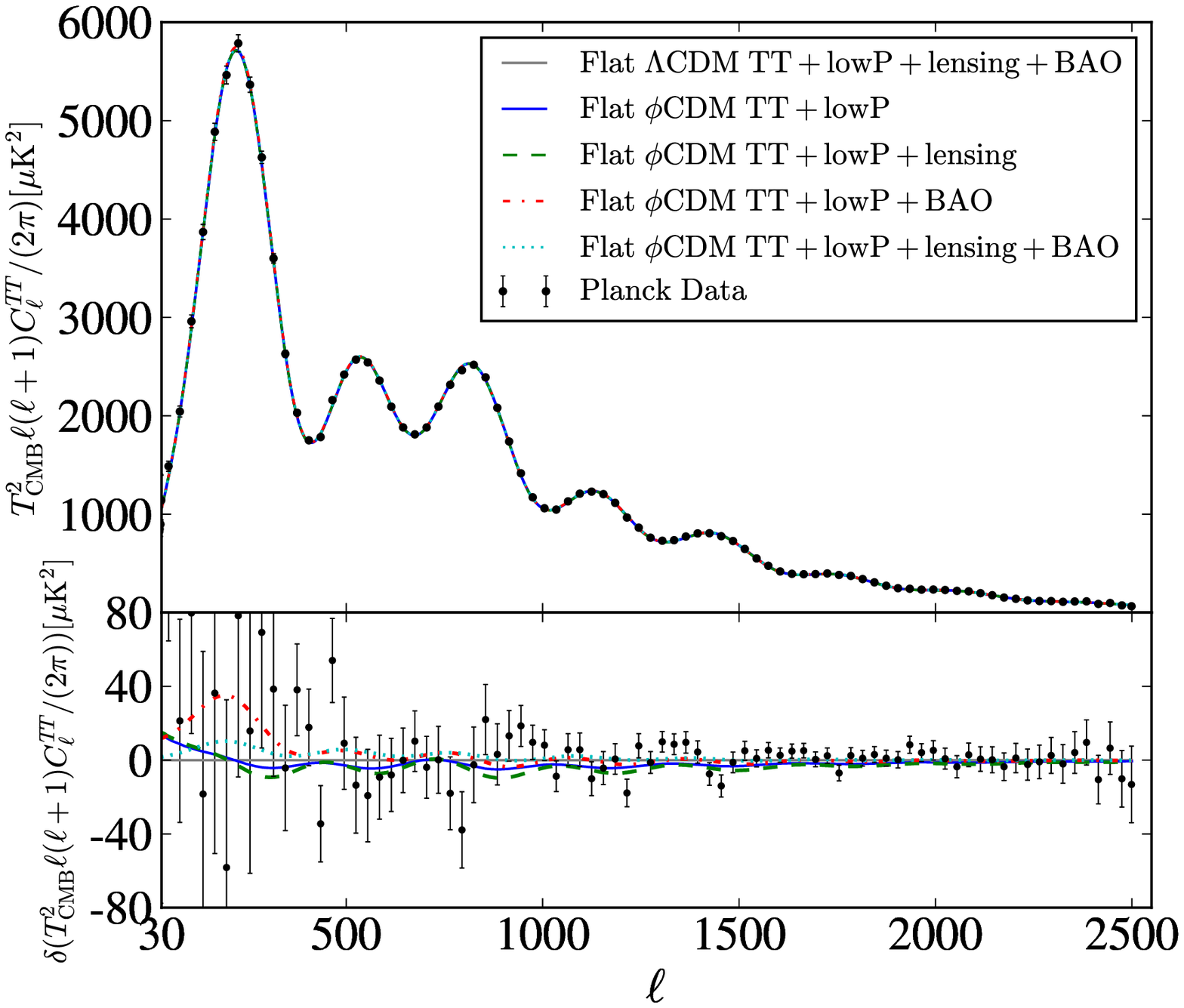}{0.43\textwidth}{(f)}
          }
\caption{The best-fit $C_{\ell}$'s for the XCDM parameterization (left panels 
(a), (c) and (e)) and the $\phi$CDM model (right panels (b), (d) and (f))
compared to the spatially-flat tilted $\Lambda$CDM model (gray solid line).
Linestyle information are in the boxes in the two lowest panels.
Planck 2015 data are shown as black points with error bars.
The top panels show the all-$\ell$ region.
The middle panels show the low-$\ell$ region $C_\ell$ and residuals.
The bottom panels show the high-$\ell$ region $C_\ell$ and residuals.\label{fig:cls}}
\end{figure}

\begin{figure}[ht]
\plottwo{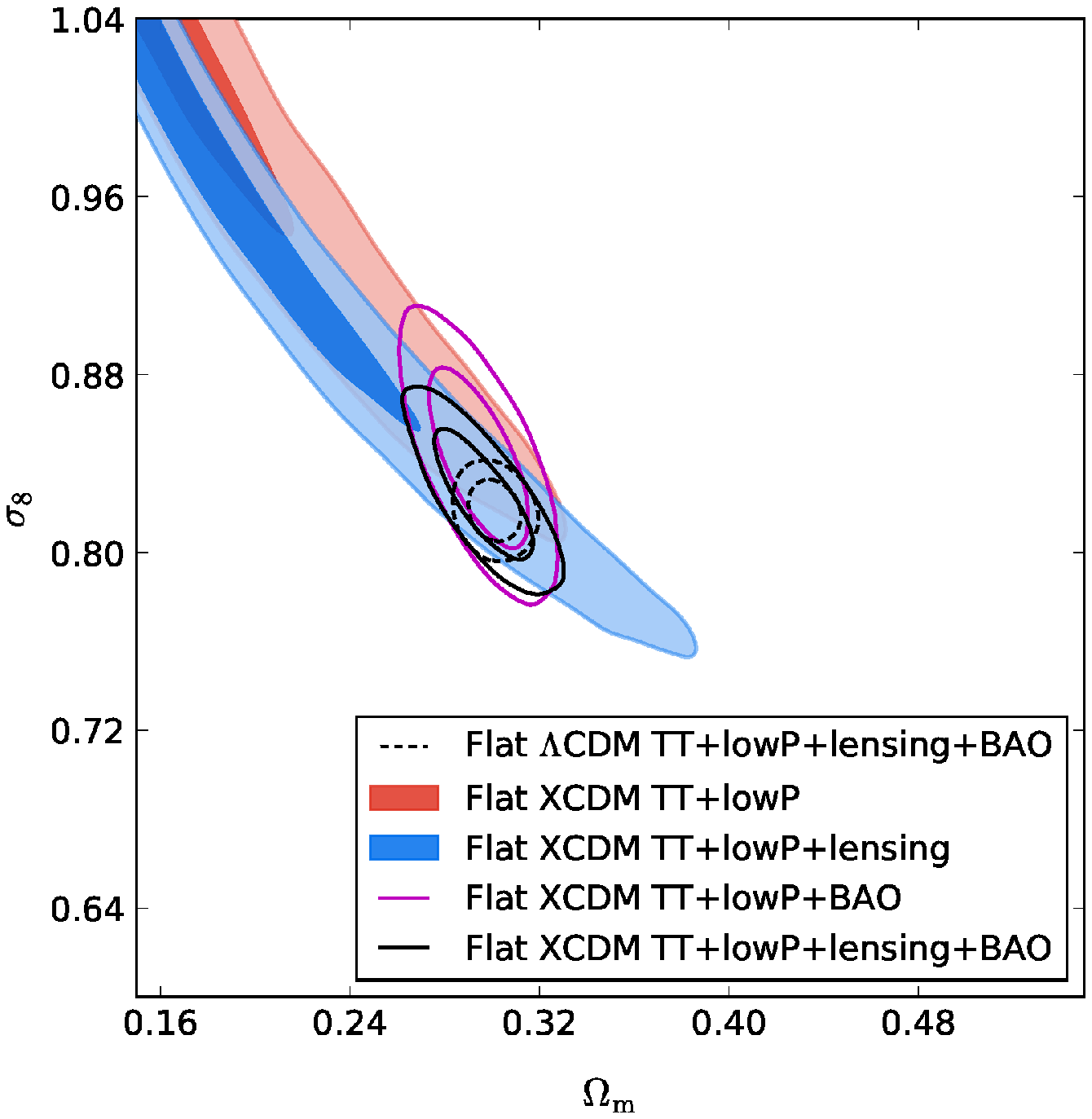}{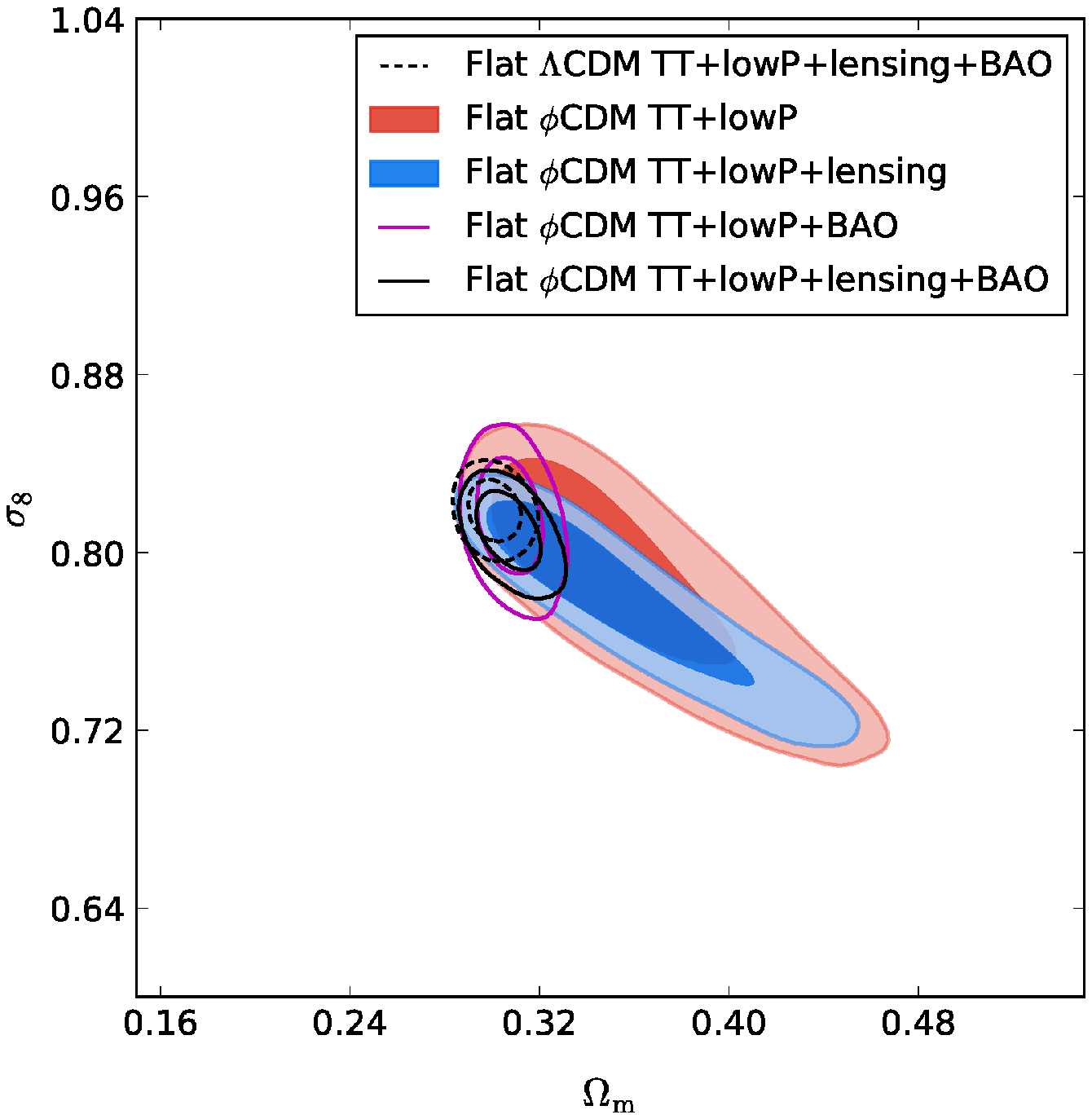}
\caption{$68.27\%$ and $95.45\%$ confidence level contours in 
the $\sigma_8$--$\Omega_{\rm m}$ plane.\label{fig:sigm}}
\end{figure}

In this section we tabulate, plot, and discuss the resulting constraints 
on the spatially-flat tilted XCDM and $\phi$CDM inflation models. 
Table \ref{tab:table1} lists mean values and $68.27\%$ limits on the 
cosmological parameters for the XCDM parameterization, and 
Table \ref{tab:table2} 
lists those for the $\phi$CDM model ($95.45\%$ upper limits on 
$\alpha$).\footnote{We thank C.-G.\ Park for pointing out a numerical error 
in our initial CMB only $\phi$CDM analyses. Our corrected 
results here are in very good agreement with those of \citet{ParkRatra2018c}.} 
Figure \ref{fig:tri} shows two-dimensional constraint contours and 
one-dimensional likelihoods from the 4 different CMB and BAO data set 
combinations used in this study. Here all other parameters are marginalized. 
CMB temperature anisotropy spectra for the best-fit XCDM and $\phi$CDM models 
are shown in Fig.\ \ref{fig:cls}, compared to that of the standard 
spatially-flat 
tilted $\Lambda$CDM model. Contours at $68.27\%$ and $95.45\%$ confidence 
level in the $\sigma_8$--$\Omega_{\rm m}$ plane are shown in Fig.\ 
\ref{fig:sigm}, with other parameters marginalized.

\begin{table*}[ht]
\caption{\label{tab:table1}
68.27\% (95.45\% on an $H_0$) confidence limits on cosmological parameters of the XCDM parameterization from CMB and BAO data.}
\centering
\begin{tabular}{lcccc}
\hline
\hline
\textrm{Parameter}&
\textrm{TT+lowP}&
\textrm{TT+lowP+lensing}&
\textrm{TT+lowP+BAO}&
\textrm{TT+lowP+lensing+BAO}\\
\hline
$\Omega_{\rm b}h^2$ & $0.02231\pm 0.00024$ & $0.02229\pm 0.00024$ & $0.02227\pm 0.00022$ & $0.02227\pm 0.00022$\\
$\Omega_{\rm c}h^2$         & $0.1191\pm 0.0023$ & $0.1182\pm 0.0021$ & $0.1189\pm 0.0019$ & $0.1183\pm 0.0018$\\
$100\theta$                 & $1.04195\pm 0.00046$ & $1.04210\pm 0.00045$ & $1.04196\pm 0.00044$ & $1.04205\pm 0.00042$\\
$\tau$                                          & $0.076\pm 0.020$ & $0.060\pm 0.018$ & $0.080\pm 0.020$ & $0.071\pm 0.017$\\
${\rm ln}(10^{10}A_{\rm s})$       & $3.085\pm 0.038$ & $3.049\pm 0.032$ & $3.092\pm 0.038$ & $3.071\pm 0.031$\\
$n_{\rm s}$        & $0.9662\pm 0.0065$ & $0.9678^{+ 0.0060}_{- 0.0064}$ & $0.9669\pm 0.0058$ & $0.9679\pm 0.0056$\\
$w_0$                   & $-1.95^{+ 0.31}_{- 0.61}$ & $-1.77^{+ 0.42}_{- 0.66}$ & $-1.06\pm 0.08$ & $-1.03\pm 0.07$\\
\hline
$H_0$ [km/s/Mpc] & $>71.26\ [2\sigma\ {\rm limit}]$ & $96.08\pm 21.63$ & $69.35\pm 1.84$ & $68.91\pm 1.78$\\
$\Omega_{\rm m}$ & $0.150^{+0.014}_{- 0.069}$ & $0.172^{+0.023}_{- 0.092}$ & $0.294\pm 0.014$ & $0.296\pm 0.014$\\
$\sigma_8$ & $1.096^{+ 0.17}_{- 0.090}$ & $1.025^{+ 0.19}_{- 0.11}$ & $0.843\pm 0.027$ & $0.826\pm 0.020$\\
\hline
\hline
\end{tabular}
\end{table*}

Our Table \ref{tab:table1} column 2 and 3 results for XCDM are in good 
agreement with the Planck 2015 results in Sec.\ 6.3 of 
\citet{PlanckCollaboration2016} and Tables 21.1 and 21.3 of ``Planck 2015 
Results: Cosmological Parameter Tables'' at 
wiki.cosmos.esa.int/planckpla2015/images/f/f7/Baseline{\_}params{\_}table{\_}2015{\_}limit68.pdf for most variables (and our Table \ref{tab:table1} column 4 and 5 results for XCDM are in good agreement with those of Tables 21.20 and 21.21 of this compilation). However our $w_0$ (and derived $H_0$, 
$\Omega_{\rm m}$, and $\sigma_8$) values differ somewhat from the Planck 2015 
ones because of the different $H_0$ flat prior ranges used (we use 
$0.2 \le h \le 1.3$ here while Planck 2015 used $h \le 1$). 

Comparing the Table \ref{tab:table1} column 5 TT + lowP + lensing +BAO results
for the spatially-flat tilted XCDM model here to those for the non-flat XCDM
model in column 5 of Table 1 in \citet{Oobaetal2018a}, we see that 
$\Omega_{\rm c} h^2$, 
$\Omega_{\rm b} h^2$, $\tau$, $\ln (10^{10} A_{\rm s})$, $\theta$, $\Omega_{\rm m}$, $w_0$, 
$H_0$, and $\sigma_8$ differ by 4.2$\sigma$, 2.5$\sigma$, 2.2$\sigma$, 
1.8$\sigma$, 1.6$\sigma$, 0.66$\sigma$, 0.25$\sigma$, 0.21$\sigma$, and 
0.20$\sigma$ (of the quadrature sum of the two error bars). Similarly for the 
$\phi$CDM case in column 5 of Table \ref{tab:table2} here and Table 2 of 
\citet{Oobaetal2018c}, we find that $\Omega_{\rm c} h^2$, $\tau$, $\Omega_{\rm b} h^2$, 
$\ln (10^{10} A_{\rm s})$, $\theta$, $\Omega_{\rm m}$, $H_0$, and $\sigma_8$ differ by 
4.2$\sigma$, 2.4$\sigma$, 2.1$\sigma$, 2.0$\sigma$, 1.4$\sigma$, 1.3$\sigma$, 
0.17$\sigma$, and 0.16$\sigma$ (of the quadrature sum of the two error bars).
On the other hand, comparing the spatially-flat tilted XCDM and $\phi$CDM
TT + lowP + lensing + BAO results we have derived here (and listed in columns 
5 of Tables \ref{tab:table1} and \ref{tab:table2}) we see that $H_0$, 
$\sigma_8$, $\Omega_{\rm m}$, $\Omega_{\rm c} h^2$, $\ln (10^{10} A_{\rm s})$, $n_{\rm s}$, $\tau$,
$\Omega_{\rm b} h^2$, and $\theta$ differ by 0.79$\sigma$, 0.73$\sigma$, 0.66$\sigma$, 0.61$\sigma$, 
0.50$\sigma$, 0.49$\sigma$, 0.49$\sigma$, 0.36$\sigma$, and 0.24$\sigma$
(of the quadrature sum of the two error bars). We note however that 
XCDM is not a physical model and so it might not be very 
meaningful to compare cosmological parameter values measured using the XCDM 
parameterization and the $\phi$CDM model. 

In agreement with \citet{ParkRatra2018a}, who compared cosmological parameter 
measurements made from cosmological observations by using the spatially-flat
tilted $\Lambda$CDM model and the non-flat $\Lambda$CDM model, we also find 
that when space curvature is allowed to vary many cosmological parameters 
cannot be determined in a model independent way from cosmological data, 
with the possible exceptions of $\sigma_8$ and $H_0$ (and $w_0$ in the XCDM
parameterization). We emphasize that the somewhat widely held belief that 
the baryonic matter density $\Omega_{\rm b} h^2$ can be pinned down in a model 
independent manner by CMB anisotropy and other cosmological observations is 
not true.\footnote{See \citet{Pentonetal2018} for a discussion of how observed deuterium abundances can be used to constrain spatial curvature.} 
When spatial curvature vanishes it appears that cosmological 
parameters can be determined in a more model independent fashion, although,
again, this is based on using the somewhat arbitrary XCDM 
parameterization. It is interesting that in this case $H_0$ and $\sigma_8$ 
are the most model dependent parameters.

\begin{table*}[ht]
\caption{\label{tab:table2}
68.27\% (95.45\% on $\alpha$) confidence limits on cosmological parameters of the $\phi$CDM model from CMB and BAO data.}
\centering
\begin{tabular}{lcccc}
\hline
\hline
\textrm{Parameter}&
\textrm{TT+lowP}&
\textrm{TT+lowP+lensing}&
\textrm{TT+lowP+BAO}&
\textrm{TT+lowP+lensing+BAO}\\
\hline
$\Omega_{\rm b}h^2$ & $0.02218\pm 0.00024$ & $0.02220\pm 0.00024$ & $0.02239\pm 0.00021$ & $0.02238\pm 0.00021$\\
$\Omega_{\rm c}h^2$         & $0.1199\pm 0.0023$ & $0.1192\pm 0.0021$ & $0.1171\pm 0.0015$ & $0.1169\pm 0.0014$\\
$100\theta$                 & $1.04184\pm 0.00045$ & $1.04193\pm 0.00044$ & $1.04215\pm 0.00042$ & $1.04219\pm 0.00041$\\
$\tau$                                          & $0.077\pm 0.019$ & $0.073\pm 0.017$ & $0.088\pm 0.019$ & $0.082\pm 0.015$\\
${\rm ln}(10^{10}A_{\rm s})$       & $3.089\pm 0.037$ & $3.078\pm 0.030$ & $3.104\pm 0.037$ & $3.092\pm 0.028$\\
$n_{\rm s}$                          & $0.9643\pm 0.0064$ & $0.9657\pm 0.0060$ & $0.9714\pm 0.0051$ & $0.9716\pm 0.0050$\\
$\alpha$\ $[2\sigma\ {\rm limit}]$                                 & $<1.46$ & $<1.19$ & $<0.28$ & $<0.28$\\
\hline
$H_0$ [km/s/Mpc] & $63.37\pm 3.00$ & $63.69\pm 3.11$ & $67.32\pm 0.89$ & $67.33\pm 0.90$\\
$\Omega_{\rm m}$ & $0.357\pm 0.035$ & $0.352\pm 0.036$ & $0.308\pm 0.009$ & $0.307\pm 0.009$\\
$\sigma_8$ & $0.789\pm 0.031$ & $0.783\pm 0.028$ & $0.815\pm 0.018$ & $0.809\pm 0.012$\\
\hline
\hline
\end{tabular}
\end{table*}

Focusing again on the TT + lowP + lensing + BAO data, we measure
$H_0 = 68.91 \pm 1.78 \  (67.33 \pm 0.90)$ km s${}^{-1}$ Mpc${}^{-1}$ for 
XCDM ($\phi$CDM), both of which are consistent with the most recent median 
statistics estimate of $H_0=68 \pm 2.8$ km s$^{-1}$ Mpc$^{-1}$ 
\citep{ChenRatra2011a},  They also are consistent with many other 
recent estimates \citep{Calabreseetal2012, Sieversetal2013, Aubourgetal2015, Chenetal2017, LinIshak2017, DESCollaboration2017b, Yuetal2018, Haridasuetal2018}, although
both are lower than the recent local expansion rate determination of 
$H_0=73.45 \pm 1.66$ km s$^{-1}$ Mpc$^{-1}$ \citep{Riessetal2018}.
 
The TT + lowP + lensing + BAO values of 
$\tau = 0.071 \pm 0.017 \  (0.082 \pm 0.015)$ for XCDM ($\phi$CDM)
measured here are a bit larger than the value of $\tau = 0.066 \pm 0.013$
measured using TT + lowP + lensing + New BAO data in the tilted 
flat-$\Lambda$CDM model \citep{ParkRatra2018a}, but not as large as
the values of $\tau$ found in the non-flat models, for 
TT + lowP + lensing + NewBAO in non-flat $\Lambda$CDM $\tau = 0.115 \pm 0.011$ 
\citep{ParkRatra2018a}, and for TT + lowP + lensing + BAO in non-flat
XCDM ($\phi$CDM) $\tau = 0.121 \pm 0.015 \  (0.129 \pm 0.013)$
\citep[][Tables 1]{Oobaetal2018b, Oobaetal2018c}. The larger value
for $\tau$ in the non-flat $\Lambda$CDM case has very interesting implications 
for reionization \citep{Mitraetal2018, Mitraetal2019}.

In both dynamical dark energy models, XCDM and $\phi$CDM, the data favor
non-evolving dark energy, although they are not yet good enough to rule 
out the possibility of mild dark energy time evolution. More and 
better-quality data will be needed to resolve this issue. The situation in the 
non-flat models is quite different, where the data favor mildly closed 
models in which the curvature energy density contributes about a per cent 
to the current cosmological energy budget \citep{Oobaetal2018a, Oobaetal2018b, Oobaetal2018c, ParkRatra2018a, ParkRatra2018b, ParkRatra2018c, ParkRatra2019d}, at 5.2$\sigma$ significance in the non-flat 
$\Lambda$CDM case for the biggest compilation of reliable cosmological 
observations \citep{ParkRatra2018b}.

The Dark Energy Survey \citep{DESCollaboration2017a} measures 
$\Omega_{\rm m}=0.264_{-0.019}^{+0.032}$ and $\sigma_8=0.807_{-0.041}^{+0.062}$
(DES Y1 All, both 68.27\% confidence limits). Our XCDM and $\phi$CDM
TT + lowP + lensing + BAO results are consistent with these limits
(with our $\phi$CDM $\Omega_{\rm m}$ value being the most deviant,
high by 1.3$\sigma$ of the quadrature sum of the two error bars).
The Dark Energy Survey constraints are also consistent with the 
XCDM and $\phi$CDM confidence level contours in the 
$\sigma_8$--$\Omega_{\rm m}$ plane shown in Fig.\ \ref{fig:sigm}, but 
are a little more difficult to reconcile with the standard 
tilted flat-$\Lambda$CDM model results. \citet{GomezValentSola2017} draw a 
similar conclusion for these models. The non-flat models are also 
more consistent with the weak lensing constraints \citep{Oobaetal2018a, Oobaetal2018b, Oobaetal2018c, ParkRatra2018a, ParkRatra2018b, ParkRatra2018c, ParkRatra2019d} than is the standard $\Lambda$CDM model.

As can be seen from  Fig.\ \ref{fig:cls} here, and the corresponding ones 
for the non-flat models \citep{Oobaetal2018a, Oobaetal2018b, Oobaetal2018c, ParkRatra2018a, ParkRatra2018b, ParkRatra2018c}, the spatially-flat tilted XCDM 
and $\phi$CDM models do not do as well at fitting the lower-$\ell$ $C_\ell$ 
temperature data as do the non-flat 
models, but the flat models better fit the higher-$\ell$ $C_\ell$'s than
do the non-flat ones.

While both spatially-flat dynamical dark energy models considered here 
are more consistent with the weak lensing constraints than is tilted 
flat-$\Lambda$CDM, the XCDM parameterization and the $\phi$CDM model 
both have one extra parameter so it is necessary to quantify how 
well these models fit the totality of 
data. Table \ref{tab:table3} shows $\Delta \chi^2_{\rm eff}$ values for 
the spatially-flat XCDM and $\phi$CDM models relative to the flat-$\Lambda$CDM 
model. Here $\chi^2_{\rm eff}$ is determined from the maximum value of the 
likelihood, $\chi^2_{\rm eff} = -2 {\rm ln} ({\rm L_{\rm max}})$.
Unlike the non-flat $\Lambda$CDM, XCDM, and $\phi$CDM models which 
are not straightforwardly related to the standard $\Lambda$CDM model
\citep{Oobaetal2018a, Oobaetal2018b, Oobaetal2018c, ParkRatra2018a, ParkRatra2018b, ParkRatra2018c}, 
the tilted spatially-flat XCDM and $\phi$CDM models here are single parameter 
extensions of the tilted flat-$\Lambda$CDM model and so we are comparing 
nested models here. In this case we can work around the ambiguity in the 
number of Planck 2015 data points and translate the 
$\Delta \chi^2_{\rm eff}$ values of Table \ref{tab:table3} to relative 
probabilities. From Table \ref{tab:table3}, for the TT + lowP + lensing + BAO 
case, the XCDM parameterization and the $\phi$CDM model, from 
$\sqrt{-\Delta \chi^2_{\rm eff}}$ for one additional free parameter, 
are 1.1$\sigma$ and 1.3$\sigma$ better fits to the data than is tilted 
flat-$\Lambda$CDM.\footnote{Closed-$\phi$CDM is also 
the best fitting of the three closed models when BAO data is 
included \citep{Oobaetal2018c}.}
The corresponding $p$ values are 0.26 and 0.21 with one additional degree of
freedom. These results indicate that the improvement in fit, in going 
from tilted flat-$\Lambda$CDM to one of the tilted spatially-flat dynamical
dark energy models, is not significant. On the other hand, the dynamical
dark energy models cannot be ruled out and continue to be of interest, 
especially $\phi$CDM which is a physically consistent model. This means that more and better-quality data will be needed to determine whether the 
dark energy density is constant or decreases slowly with time. For 
goodness-of-fit, one may also 
consider the AIC, which will penalize the dynamical dark energy models for 
the additional parameter. From $\Delta$AIC the XCDM ($\phi$CDM) model is 
only 69\% (82\%) as likely as the standard tilted flat-$\Lambda$CDM model, 
again not a strong result either way.

\begin{table*}[ht]
\caption{\label{tab:table3}
$\Delta \chi^2_{\rm eff}$ values for the best-fit XCDM ($\phi$CDM) model.}
\centering
\begin{tabular}{lcc}
\hline
\hline
\textrm{Data sets}& \textrm{$\Delta \chi^2_{\rm eff}$}\\
\hline
TT+lowP & $-3.27\ (+0.41)$\\
TT+lowP+lensing & $-1.89\ (-0.79)$\\
TT+lowP+BAO & $-0.82\ (-0.64)$\\
TT+lowP+lensing+BAO & $-1.26\ (-1.60)$\\
\hline
\hline
\end{tabular}
\end{table*}

\section{Conclusion}

We present constraints on the tilted spatially-flat XCDM and $\phi$CDM 
inflation models determined by analyzing Planck 2015 CMB anisotropy data
as well as BAO distance measurements. XCDM is a simply parameterized 
dynamical dark energy model, and $\phi$CDM is a physically consistent one 
in which a scalar field $\phi$ with an inverse power-law potential energy
density acts as dynamical dark energy and powers the currently 
accelerating cosmological expansion. Both of these dynamical dark energy
models better fit, although not significantly so, the TT + lowP + lensing + 
BAO data combination than does the tilted flat-$\Lambda$CDM model. Perhaps
more interestingly, the dynamical dark energy models reduce the tension 
between the Planck 2015 CMB anisotropy and the weak lensing $\sigma_8$ 
constraints. More and better data, which should soon be available, is needed
to determine if dynamical dark energy can be ruled out, or if dark energy is dynamical.

\section*{Acknowledgments}

We thank G.\ Horton-Smith and C.-G.\ Park for helpful discussions. We 
thank the referee for comments that helped us improve the paper.
This work is supported by Grants-in-Aid for Scientific Research 
from JSPS (Nos.\ 16J05446 (J.O.) and 15H05890 (N.S.)). B.R.\ is supported 
in part by DOE grant DE-SC0019038.


\end{document}